\documentclass{article}
\usepackage{dcase2023,amsmath,graphicx,url,times,booktabs, tabularx, enumitem}
\usepackage{textcomp}
\usepackage{xcolor}
\usepackage{tikz}
\usepackage{pgfplots}
\usepackage{scalerel}
\usepackage{color}
\usepackage{comment}
\usepackage{scalerel}
\usepackage{upgreek}
\usepackage{multirow}
\usepackage{makecell}
\usepackage{amssymb}
\usepackage{hhline}
\usepackage{pifont}
\usepackage{cite}
\newcommand{\cmark}{\ding{51}}%
\newcommand{\xmark}{\ding{55}}%

\usepackage[export]{adjustbox}
\def \am [#1]{\textcolor{red}{AM: #1}}
\let\lctau\tau % save the lowercase of '\tau'
\renewcommand{\tau}{\scalerel*{\lctau}{X}}
\def\BibTeX{{\rm B\kern-.05em{\sc i\kern-.025em b}\kern-.08em
    T\kern-.1667em\lower.7ex\hbox{E}\kern-.125emX}}

\title{Incremental Learning of Acoustic Scenes and Sound Events}

 \name{Manjunath Mulimani, Annamaria Mesaros\thanks{This work was supported in part by Academy of Finland grant 332063 ``Teaching machines to listen". The authors wish to thank CSC-IT Centre of Science Ltd., Finland,  for providing computational resources.}}
 \address{ Computing Sciences, Tampere University,
Tampere, Finland\\
\{manjunath.mulimani, annamaria.mesaros\}@tuni.fi}

\begin{document}

\ninept
\maketitle

\begin{sloppy}

\begin{abstract}
In this paper, we propose a method for incremental learning of two distinct tasks over time: acoustic scene classification (ASC) and audio tagging (AT). We use a simple convolutional neural network (CNN) model as an incremental learner to solve the tasks. Generally, incremental learning methods catastrophically forget the previous task when sequentially trained on a new task. To alleviate this problem, we propose independent learning and knowledge distillation (KD) between the timesteps in learning.
Experiments are performed on TUT 2016/2017 dataset, containing  4 acoustic scene classes and 25 sound event classes. 
The proposed incremental learner first solves the ASC task with an accuracy of 94.0\%. Next, it learns to solve the AT task with an F1 score of 54.4\%. At the same time, its performance on the previous ASC task decreases only by 5.1 percentage points due to the additional learning of the AT task. 
\end{abstract}

\begin{keywords}
Incremental learning,  independent learning, knowledge distillation, acoustic scene classification, audio tagging
\end{keywords}

\section{Introduction}
\label{sec:intro}
The natural learning system of humans incrementally learns new concepts over time without forgetting the previously learned ones.  This process of learning is known as continuous, incremental, or lifelong learning. In contrast, deep learning-based systems have the ability to learn a task efficiently, but fine-tuning the same system with a new task tends to override the previously acquired knowledge. This leads to a phenomenon of deteriorating performance on previously learned tasks known as catastrophic forgetting. Developing a robust system that should not degrade its performance significantly on previous tasks as new tasks are added is currently a pursued problem in many domains.

\begin{figure}
    \centering
    \includegraphics[width=\linewidth]{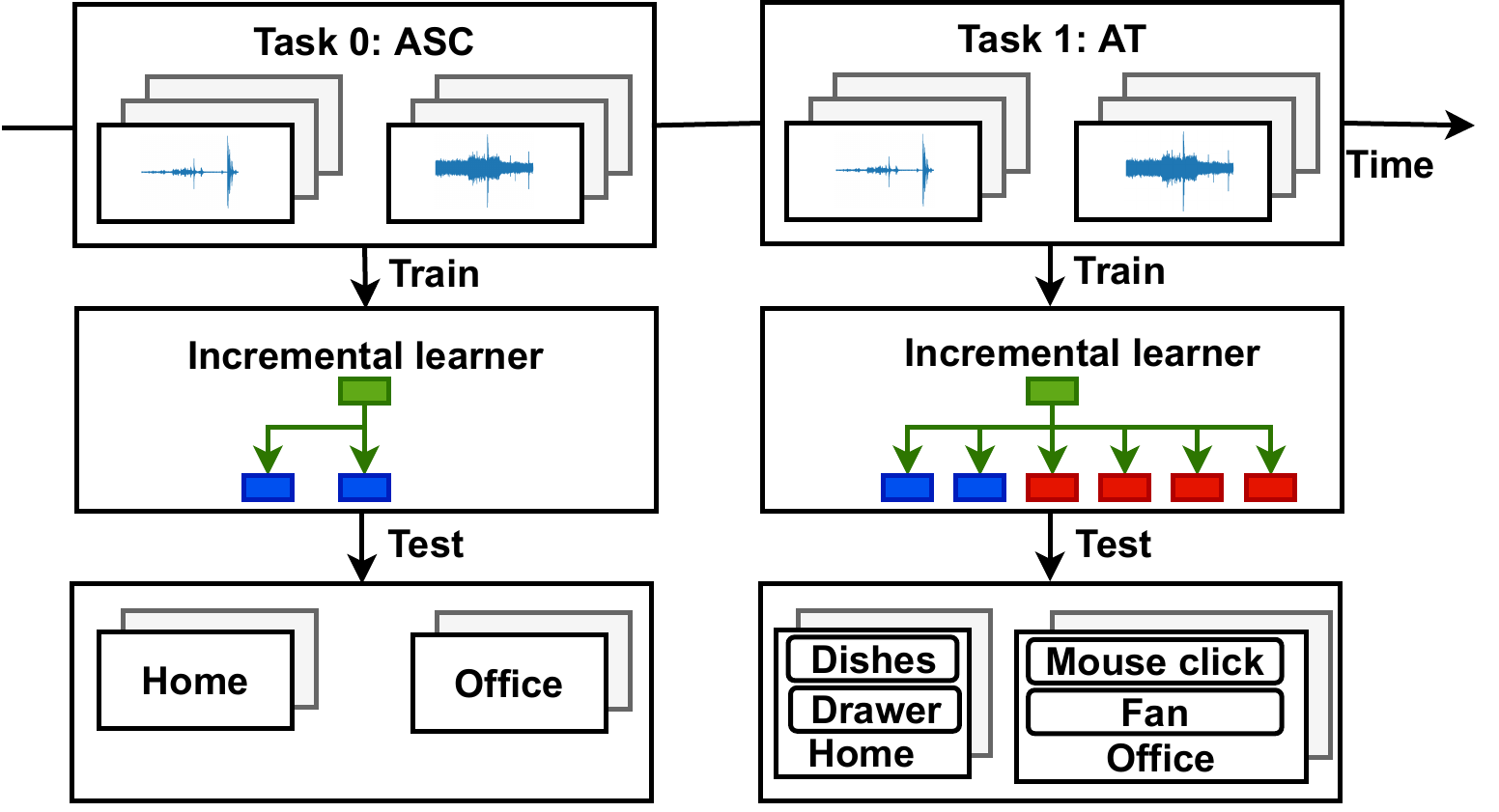}
    \caption{Incremental learning of distinct tasks: acoustic scene classification (ASC) and audio tagging (AT). Our incremental learner learns acoustic scene classes initially (blue units) and sound event classes incrementally (red units). After the learning process of each task, a learner is evaluated on classes of all tasks learned so far.
}
    \label{fig:block_diagram}
    \vspace{-10pt}
\end{figure}

Most of the studies reported in the literature on incremental learning operate on images, e.g. object detection \cite{kj2021incremental, perez2020incremental}, image classification \cite{mittal2021essentials, ahn2021ss}, and semantic segmentation \cite{yu2022self, cermelli2022incremental}. 
A few works report on incremental learning of audio such as environmental sound classification (ESC) \cite{wang2021few, bayram2021incremental},  audio captioning \cite{berg2021continual}, and fake audio detection \cite{ma21b_interspeech}. 
However, these methods are restricted to solving an initial base task followed by $N$ incremental tasks of the same problem (e.g. ESC). In addition, these methods were designed to have the same number of classes in the incremental tasks, an assumption that does not hold in practical scenarios. Furthermore, most of these methods use a small portion of data from the previous task during the training of the system on current task data and employ complex postprocessing methods to alleviate catastrophic forgetting \cite{joseph2022energy}.

In this work, we propose an incremental learning method to solve distinct tasks over time: acoustic scene classification (ASC) and audio tagging (AT) (see Fig.~\ref{fig:block_diagram}), to simulate the scenario of new information becoming available at a later time for the same audio material. In this case, the incremental learner first learns the acoustic scenes and then learns more detailed characterization of the acoustic content of the scene, i.e. the sound events active in the given acoustic scene. When learning the sound event classes, the same audio material is used, but the acoustic scene labels are no longer available in the learning process.

The proposed work is based on the class incremental approach \cite{rebuffi2017icarl} and uses a single classifier to learn both acoustic scenes and sound events rather than using a separate classifier for each task, as done in \cite{li2017learning}. Using the same classifier makes the problem more challenging, because acoustic scene and sound event classes compete in a single classifier. In comparison \cite{li2017learning} employs information on the task identity to get output from a specific classifier at the prediction step. 
Inspired by different incremental learning methods for image classification \cite{li2017learning, rebuffi2017icarl, hou2019learning, ahn2021ss, mittal2021essentials}, we propose a simple CNN-based incremental learner to solve distinct tasks over time.  

The main contributions of this work are summarized below.
\begin{itemize}
[noitemsep,nolistsep,leftmargin=*]  
    \item We design an independent learning (IndL) mechanism that allows a classifier to learn different tasks effectively.
    \item We combine IndL with  Kullback-Leibler (KL) divergence-based distillation loss to learn new sound events while preserving the knowledge of acoustic scene classes.
    \item We conduct experiments on both two-step ASC-AT and three-step ASC-ASC-AT incremental setups to analyse the behavior of the proposed incremental learner over time.
\end{itemize}
  
The rest of the paper is organized as follows. Section 2 introduces the proposed framework for incremental learning of acoustic scenes and sound events tasks. Section 3 presents the different experiments that compare incremental learning with multi-task learning and separate models for the same tasks. Finally, conclusions and future work are given in Section 4.

\section{Incremental learning}

In our incremental learning setting, a set of ASC and AT tasks $\{\lctau_0, \lctau_1,\ldots,\lctau_t\}$ is presented to a learner sequentially in incremental time steps $t$. The isolated task 
$\lctau_t=\{(\mathbf{x}_i^{\lctau_t}, \mathbf{y}_i^{\lctau_t})|1\leq i\leq m\}$
presented at time step $t$ is composed of input features $\mathbf{x}_i^{\lctau_t}$ and corresponding one-hot (for ASC) or multi-hot (for AT) ground truth label vectors $\mathbf{y}_i^{\lctau_t}\in \{0, 1\}^{C_t}$. $C_t$ denotes the number of classes in tasks up to and including task ${\lctau_t}$. 
The two distinct ASC and AT tasks use the same audio clips but none of the tasks share the class labels. Specifically, the learner does not have access to labels of the $\lctau_0$ while learning the $\lctau_1$. 
Typically, this class imbalance makes the learner's predictions biased to focus on the classes in the current task and catastrophically forget the classes of the previous task.

In this work, our goal is to build a learner $\mathcal{P}^{\lctau_t}$, which can solve all the tasks learned so far. 
A learner $\mathcal{P}^{\lctau_t}$ is a deep network that includes a feature extractor $\mathcal{F}_\mathbf{\theta}^{\lctau_t}$ (parameterized by weights $\theta$) and a fully connected layer $\mathcal{F}_\phi^{\lctau_t}$ (parameterized by weights $\phi$) for classification. %of acoustic scenes or tagging of acoustic events. 
Output logits of the network on a given input $\mathbf{x}$ are obtained by $\mathbf{o}(\mathbf{x}) = \mathcal{F}_\phi^{\lctau_t}(\mathcal{F}_\theta^{\lctau_t}(\mathbf{x}))$. 

Generally, incremental learners learn to solve one initial base task followed by similar incremental tasks, for example like the setup in \cite{wang2021few} where the initial task was learning 30 sound classes, then learning sets of 10 sound classes in incremental time steps for the same classification task. However, in this work, we solve distinct incremental tasks: single-label ASC and multi-label AT. We experiment with two scenarios. In the first scenario, the learner solves ASC in an initial step ($\lctau_0=\text{ASC}$) and AT in the next time step ($\lctau_1=\text{AT}$) as depicted in Fig.~\ref{fig:block_diagram} (hereafter referred to as ASC-AT task). In the second scenario, the learner solves ASC in an initial step ($\lctau_0=\text{ASC}$), and  ASC and AT in subsequent incremental time steps ($\lctau_1=\text{ASC}$ and $\lctau_2=\text{AT}$, hereafter referred to as ASC-ASC-AT). 
\subsection{Incremental ASC-AT learning}
 \label{n1}
In the initial time step $t=0$,  a learner  $\mathcal{P}^{\lctau_0}$ learns $\mathcal{F}_\phi^{\lctau_0}$ to classify the acoustic scene classes of a task $\lctau_0$.   $\mathcal{P}^{\lctau_0}$ is trained using cross-entropy loss $ \mathcal{L}^{CE}$ computed using softmax $\sigma$ over logits $\mathbf{o}$ as: 
\begin{equation}
    \mathcal{L}^{CE} = -\sum_{k=1}^{C}\mathbf{y}_k^{\lctau_0}\cdot\log(\sigma(\mathbf{o}_k)),
    \label{eq1}
\end{equation}

In the incremental time step $t=1$, a new learner $\mathcal{P}^{\lctau_{1}}$ is initialized by previous learner $\mathcal{P}^{\lctau_{0}}$. The classifier $\mathcal{F}_\phi^{\lctau_{1}}$ of  $\mathcal{P}^{\lctau_{1}}$ is extended to learn sound event classes of a task $\lctau_{1}$ by adding new output units. The output logits of $\mathcal{P}^{\lctau_{1}}$ comprise $\mathbf{o}=\{\mathbf{o}^{old}, \mathbf{o}^{new}\}$. $\mathbf{o}^{old}$ and $\mathbf{o}^{new}$ denote the logits of acoustic scene classes and sound event classes respectively. We propose an independent learning process (IndL) through which  the entire $\mathcal{P}^{\lctau_{1}}$ is trained using separate losses for $\mathbf{o}^{old}$ and $\mathbf{o}^{new}$ logits (see Fig.~\ref{losses}).  

A binary cross-entropy loss $ \mathcal{L}^{BCE}$ is computed using sigmoid $\sigma$ over logits of the novel acoustic event classes $\mathbf{o}^{new}$ only:
 \begin{multline}
      \mathcal{L}^{BCE} = -\sum_{k=C_{t-1}+1}^{C_t}\mathbf{y}_k^{\lctau_{1}}\cdot\log(\sigma(\mathbf{o}_k^{new})) \\ +(1-\mathbf{y}_k^{\lctau_{1}})\cdot\log(1-\sigma(\mathbf{o}_k^{new})),
      \label{eq2}
 \end{multline}  
where $C_{t-1}$ denotes the number of classes up to and excluding task $\lctau_{1}$. 
This independent learning of weights $\phi$ of $\mathcal{F}_\phi^{\lctau_{1}}$ of the novel sound event classes from previous acoustic scene classes is meant to reduce the catastrophic forgetting. To be more specific, IndL allows $\mathcal{P}^{\lctau_{1}}$ to learn from Eq. (\ref{eq2}) using only the $\mathbf{o}^{new}$ logits of the sound event classes, % without disturbing the $\mathbf{o}^{old}$ logits of the acoustic scene classes. The 
while $\mathbf{o}^{old}$ logits of the acoustic scene classes are handled separately. This independence of the two tasks is also evident in the different loss functions per task ($\mathcal{L}^{CE}$ vs $\mathcal{L}^{BCE}$). 

A distillation loss $\mathcal{L}^{KD}$ is computed using  Kullback-Leibler divergence ($\mathcal{D}_{KL}$) between  $\mathbf{o}^{old}$ logits of current $\mathcal{P}^{\lctau_{1}}$ learner and output logits of previous frozen $\hat{\mathcal{P}}^{\lctau_0}$ learner:
\begin{equation}
    \mathcal{L}^{KD} =\mathcal{D}_{KL}(\hat{\mathbf{v}}||\mathbf{v}),
    \label{eq3}
\end{equation}
where  $\mathbf{v}=\sigma(\frac{\mathbf{o}^{old}}{T})$ denotes the $\mathbf{o}^{old}$ logits of $\mathcal{P}^{\lctau_{1}}$ and $\hat{\mathbf{v}}=\sigma(\frac{\hat{\mathcal{P}}^{\lctau_0}(\mathbf{x})}{T})$ denotes the logits of $\hat{\mathcal{P}}^{\lctau_0}$. The $\sigma$ is the softmax and $T$ is the temperature hyperparameter to smooth the $\mathcal{L}^{KD}$.
The $\mathcal{L}^{KD}$ acts as a forgetting constraint that penalizes the change concerning the output of the previous learner. Specifically, the learner $\mathcal{P}^{\lctau_{1}}$  preserves the knowledge about the previous ASC task using $\mathcal{L}^{KD}$ and continues to learn the new AT task using $\mathcal{L}^{BCE}$.
Therefore, $\mathcal{P}^{\lctau_{1}}$ is trained using combined loss as:
\begin{equation}
    \mathcal{L} = \mathcal{L}^{BCE}+\lambda\mathcal{L}^{KD},
    \label{eq4}
\end{equation}
where $\lambda$ denotes the weight of the loss which we adaptively set to $\Omega\sqrt{C_{t}-C_{t-1}/C_t}$ as per the recommendation of \cite{hou2019learning}. $\Omega$ is a constant. $C_{t}-C_{t-1}$ denotes the number of new sound event classes.
 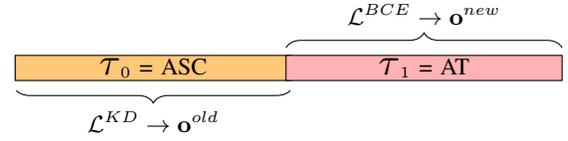
\begin{figure}
     \centering
\begin{tikzpicture}[
box/.style = {draw, minimum size=9.5pt,inner sep=1pt, outer sep=0pt, right, text width=3.6cm, align=center},
brc/.style = {decorate, decoration={brace,raise=1mm, amplitude=2mm}},
shorten <>/.style = {shorten >=#1pt, shorten <=#1pt}
                     ]
\coordinate (a0) at (0,0);
\node[box, fill={rgb:orange,1;yellow,2;pink,5}] (a1) at (0, 0) {$\tau_0$ = ASC};
\node[box, fill=red!30] (a2) at (3.6, 0) {$\tau_1$ = AT};
\draw[brc] (a2.north west) to node[above=3mm] {$\mathcal{L}^{BCE}\rightarrow\mathbf{o}^{new}$} (a2.north east);
\def\brcpad{1pt}
\draw[brc,shorten <>=1] (a1.south east) to node[below=3mm] {$\mathcal{L}^{KD}\rightarrow\mathbf{o}^{old}$} (a1.south
west);
\end{tikzpicture}
 \caption{Incremental ASC-AT learning; the different losses are calculated on logits of the previous ASC task and logits of the new AT task separately at the incremental time step $t=1$.}
     \label{losses}
     \vspace{-10pt}
\end{figure}

We use two more techniques reported in the literature for incremental learning of images. % to reduce forgetting previous classes and effectively recognize the new classes.
One is, that the learning rate (LR) is reduced in incremental time steps, as done in \cite{mittal2021essentials}. This was shown to improve the transfer of knowledge from the old to the new learner and mitigate the adverse effect of imbalanced data in incremental time steps.
Another is the use of cosine normalization in the classifier $\mathcal{F}_\phi^{\lctau_t}$ \cite{hou2019learning}. It was observed that the magnitudes of the weight and bias of the previous and current classes in $\mathcal{F}_\phi^{\lctau_t}$ are significantly different. Cosine normalization restricts the values of input distributions to $[-1, 1]$ and eliminates the bias that arises due to the magnitude difference.
\subsection{Incremental ASC-ASC-AT learning}
\label{n2}

In this case, the learner $\mathcal{P}^{\lctau_{0}}$ learns to solve an additional ASC task using  $\mathcal{L}^{CE}$ at the initial time step $t=0$.  In the incremental time step  $t=1$,  $\mathcal{P}^{\lctau_{1}}$ continues to learn a new ASC task in the absence of the old ASC task's data. Specifically, the two ASC tasks do not share audio clips nor scene classes.  %The learner uses the same audio clips of the old task only when solving the new AT task to identify sound events in acoustic scenes. 
In the incremental time step  $t=1$, $\mathcal{L}^{CE}$ is computed independently using softmax $\sigma$ over logits of the new acoustic scene classes $\mathbf{o}^{new}$ only, as per Eq. (\ref{eq1}) and  $\mathcal{L}^{KD}$ is used to hold the knowledge of old acoustic scenes, as per Eq. (\ref{eq3}). $\mathcal{P}^{\lctau_{1}}$ is trained using combined loss as:
\begin{equation}
    \mathcal{L} = \mathcal{L}^{CE}+\lambda\mathcal{L}^{KD}
    \label{eq5}
\end{equation} 
In the incremental time step $t=2$, $\mathcal{P}^{\lctau_{2}}$ learns the AT task (using $\mathcal{L}^{BCE}$)  by preserving the knowledge of all old acoustic scenes (using $\mathcal{L}^{KD}$) as per the description given in \ref{n1} and illustrated in Fig.~\ref{loss:ASC-ASC-AT}. For the AT task, the learner uses the same audio clips as the ASC task at $t=1$. 
\begin{figure}
     \centering
\begin{tikzpicture}[
box/.style = {draw, minimum size=9.5pt,inner sep=1pt, outer sep=0pt, right, text width=2.6cm, align=center},
brc/.style = {decorate, decoration={brace,raise=1mm, amplitude=2mm}},
shorten <>/.style = {shorten >=#1pt, shorten <=#1pt}
                     ]
\coordinate (a0) at (0,0);

\node[box, fill={rgb:orange,1;yellow,2;pink,5}] (a1) at (0, 0) {$\tau_0$ = ASC};
\node[box, fill={rgb:orange,1;yellow,2;pink,5}] (a2) at (2.56, 0) {$\tau_1$ = ASC};

\node[box, fill=red!30] (a3) at (5.2, 0) {$\tau_2$ = AT};
\draw[brc] (a3.north west) to node[above=3mm] {$\mathcal{L}^{BCE}\rightarrow\mathbf{o}^{new}$} (a3.north east);
\def\brcpad{1pt}

\draw[brc,shorten <>=1] (a2.south east) to node[below=3mm] {$\mathcal{L}^{KD}\rightarrow\mathbf{o}^{old}$} (a1.south
west);

\end{tikzpicture}
 \caption{Incremental ASC-ASC-AT learning; the different losses are calculated on old ASC and new AT logits separately at an incremental time step $t=2$.}
     \label{loss:ASC-ASC-AT}
     \vspace{-10pt}
\end{figure}
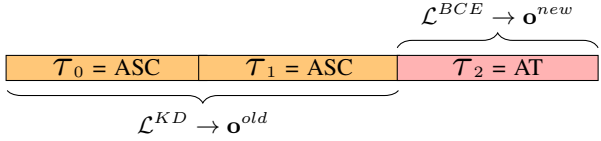

 \section{Evaluation and Results}
  \subsection{Datasets}
 For ASC-AT, 
 we use acoustic scenes and corresponding sound events from TUT 2016/2017 dataset \cite{imoto2020sound, krause2022binaural}.
 The dataset contains 192 minutes of audio recordings. Task 0 is composed of four acoustic scenes: home, residential area, city center, and office. Task 1 is composed of 25 sound events: bird singing, brakes squeaking, breathing, 
 etc. Complete details about the data can be found in \cite{imoto2020sound}.

 For ASC-ASC-AT, we use TUT Acoustic Scenes 2017 \cite{mesaros2017dcase} and TUT 2016/2017. 
 Task 0 is composed of 11 acoustic scenes: beach, bus,     cafe/restaurant, car,     forest path,     grocery store,     library,
 metro station, park, train, and tram. 
 Tasks 1 and 2 are the ASC and AT from the previous experiment. 
The learner is trained and tested on official development and evaluation splits of the datasets in each step.

\subsection{Implementation details}
\label{n4}
 Input features in each time step are 40-dimensional log mel-band energies obtained from each audio segment in 40 ms frames with 50\% overlap.
The network architecture of the feature extractor $\mathcal{F}_\mathbf{\theta}^{\mathcal{T}_t}$ 
includes three convolutional blocks, each consisting of two $3\times 3$ convolutional layers, with batch-normalization and ReLU nonlinearity applied to each convolutional layer.  $2\times 2$ average pooling is applied to each convolutional block, and 20\% dropout is applied after each average pooling to avoid overfitting. The number of feature maps of convolutional blocks is set to $\{16, 32, 64\}$. The flattened output of the last convolutional block is considered as the input to the cosine normalized fully-connected layer $\mathcal{F}_\phi^{\lctau_t}$, whose  
number of output units 
is equal to the number of classes in each time step. 

The learner's network is trained using the SGD optimizer \cite{loshchilov2017sgdr} with a momentum of 0.9 and a mini-batch size of 100 for 120 epochs. The initial learning rates for task 0 and incremental task(s) are set to 0.1 and 0.01 respectively. CosineAnnealingLR \cite{loshchilov2017sgdr} scheduler is used to update the optimizer in every epoch. Other hyper-parameters: $T$ and $\Omega$ are empirically set to 2 and 5 
 respectively. 

\begin{table}[]
\centering
\begin{tabular}{l|c|c|c|c}
\toprule
& & {$t=0$} & \multicolumn{2}{c}{$t=1$}   \\
\cline{3-5}
 Method&KD & Task 0: & Task 1: & Task 0:  \\
&& ASC (Acc) & AT (F1) & ASC (Acc) \\
\midrule 
 ASC &-& 94.0 &- & -\\
 \hline
 AT &-&- & 53.0& -\\
 \hline
 \makecell[l]{ Joint\\ ASC-AT} &-& 72.0&50.4 & -\\
 \hline
 \makecell[l]{Incremental\\ ASC-AT} &\xmark& 94.0& 54.4&84.1 (9.9$\downarrow$)\\
 \hline
 \makecell[l]{Incremental\\ ASC-AT} &\cmark&94.0 &54.4 & 88.9 (5.1$\downarrow$)\\
 \bottomrule
\end{tabular}
\caption{Incremental ASC (4 classes) and AT (25 classes) with and without KD, compared to individual and joint learning of tasks with a similar architecture. The value within () denotes the forgetting amount; $\downarrow$ indicates that lower is better. 
    }

        \label{tab:ASC-AT-10s}
\end{table}

\subsection{Baseline systems and evaluation metrics}
The performance of the proposed incremental ASC-AT %and ASC-ASC-AT 
system is compared with the individual ASC, AT, and joint ASC-AT baseline systems. The same network architecture as the incremental ASC-AT system is used in all the baseline systems for fair comparison.
Individual ASC and AT systems solve only the ASC or AT task, respectively. 
A joint ASC-AT system is a multi-task system that is trained for ASC and AT tasks at the same time using cross-entropy loss and binary cross-entropy loss respectively, as proposed in \cite{tonami2019joint}.
The performance on ASC and AT tasks is evaluated using accuracy and  F1 score (using a threshold of 0.5), respectively.

\begin{table*}[]
    \centering
  \begin{tabular}{lll|c|c|l|c|c|l}
\toprule
\multicolumn{3}{c|} {Ablation setting} & {$t=0$} & \multicolumn{2}{c|} {$t=1$}& \multicolumn{3}{c}{$t=2$}   \\
\cline{4-9}
\multicolumn{3}{c|}{} & \makecell{Task 0:}& \multicolumn{2}{c|}{Task 1: ASC (Acc)}& Task 2:  & \multicolumn{2}{c}{Task 0 and 1: ASC (Acc)} \\
%\midrule
\hhline{---~--~--}
IndL&KD& LRs &ASC (Acc) &Overall&Task-wise& AT (F1)& Overall & Task-wise \\
\midrule
\cmark&\cmark&\{0.1, 0.1\} & 65.3 &  40.0 & \makecell[l]{Task 0: 34.6 (30.7$\downarrow$)\\Task 1: 54.9} & 53.0 & 38.8 & \makecell[l]{Task 0: 32.9 (32.4$\downarrow$)\\Task 1: 54.7 (0.2$\downarrow$)}\\
\hline
\cmark&\cmark&\{0.01, 0.01\} & \textbf{68.1} &  50.1 & \makecell[l]{Task 0: 54.3 (13.8$\downarrow$)\\Task 1: 38.7} & 53.0 & 49.1 & \makecell[l]{Task 0: 53.3 (14.8$\downarrow$)\\Task 1: 37.3 (1.4$\downarrow$)}\\
\hline
\cmark&\cmark&\{0.001, 0.001\} & 49.1 &  37.2 & \makecell[l]{Task 0: 23.4 (25.7$\downarrow$)\\Task 1: 77.0} & 43.0 & 34.0 & \makecell[l]{Task 0: 17.7 (31.4$\downarrow$)\\Task 1: 75.2 (1.8$\downarrow$)}\\
\hline
\cmark&\cmark&\{0.01, 0.001\} & \textbf{68.1} &  49.3 & \makecell[l]{Task 0: 57.7 (10.4$\downarrow$)\\Task 1: 25.9} & 46.0 & 48.3 & \makecell[l]{Task 0: 56.9 (11.2$\downarrow$)\\Task 1: 24.1 (1.8$\downarrow$)}\\
\hline
\cmark&\cmark&\{0.1, 0.01\} & 65.3 &  \textbf{53.8} & \makecell[l]{Task 0:\textbf{ 54.6 }(10.7$\downarrow$)\\Task 1: \textbf{53.1}} & 53.0 & \textbf{52.2} & \makecell[l]{Task 0: \textbf{53.3} (12$\downarrow$)\\Task 1: \textbf{49.1} (4$\downarrow$)}\\
\hline
\xmark&\xmark&\{0.1, 0.01\} & 65.3 &  26.4 & \makecell[l]{Task 0: 0.1 (65.2$\downarrow$)\\ Task 1: 85.0} & 53.0& 23.2 & \makecell[l]{Task 0: 0.1 (65.2$\downarrow$) \\  Task 1: 80.4 (4.6$\downarrow$)}\\
\bottomrule
\end{tabular}
\caption{Incremental ASC-ASC-AT with and without independent learning (IndL) and knowledge distillation (KD) using different LR combinations in initial and incremental steps with Task 0: 11 classes, Task 1: 4 classes, Task 2: 25 classes.  For $t=1$ and $t=2$ the overall ASC accuracy represents performance over all 15 scene classes, and separate task-wise accuracy is provided (over the 11 classes of Task 0 and over the 4 classes of Task 1).}
        \label{tab:ASC-ASC-AT}
\vspace{-8pt}
    \end{table*}
    
\subsection{ASC-AT results}
The experimental results provided in Table \ref{tab:ASC-AT-10s} compare the performance of the proposed incremental ASC-AT system with baseline systems. 
The learning of ASC and AT tasks at the same time as joint ASC-AT results in an overall performance lower than ASC and AT learned separately in different systems. Particularly, the accuracy of the ASC side in joint ASC-AT is significantly worse. This is also true with existing ASC-sound event detection (SED) multi-task models \cite{tonami2019joint, krause2022binaural}.

On the other hand, the proposed incremental ASC-AT method can solve ASC and AT tasks with an accuracy of 84.1\% (for the system without KD) and an F1 score of 54.4\% respectively. The performance on the ASC task at the incremental time step is reduced by 9.9 percentage point (p.p.) as compared to the initial time step. Surprisingly, we see a small increase in the F1 score of the AT task as compared to the individual AT system.
We hypothesize that this is  due to the incremental learner already pre-trained using very relevant acoustic content (but different classes) in the initial time step, which may generate richer feature representations for the AT task.
We observe that using KD is more efficient for preserving the knowledge of the previous task. The incremental ASC-AT system with KD outperforms the system without KD, having an average accuracy of 88.9\% on the previous ASC task with only 5.1 p.p. reduction in performance. 

\subsection{ASC-ASC-AT results}

\begin{figure}[]
    \centering
    \includegraphics[width=\linewidth]{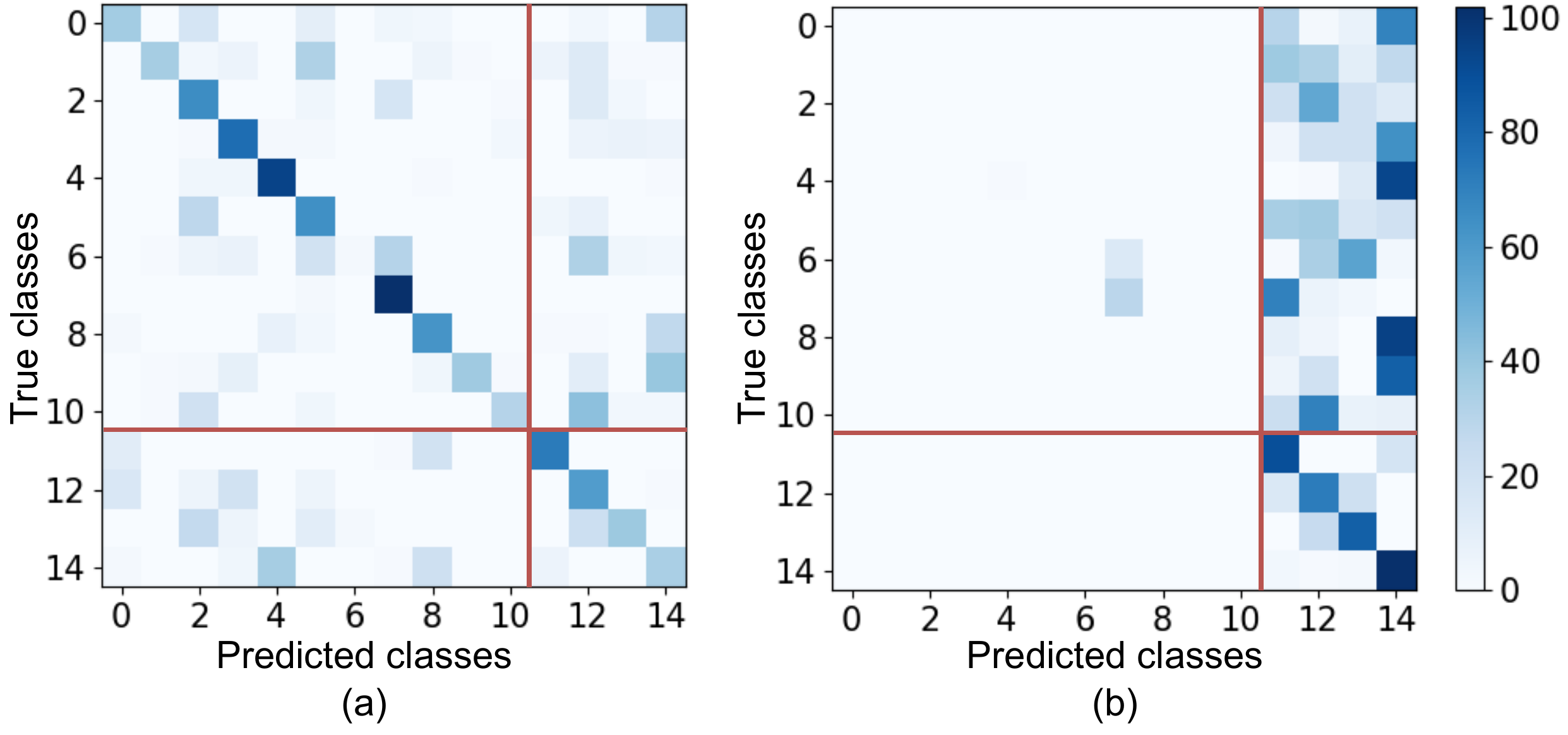}
    \caption{Confusion matrices w/ (a) and w/o (b) KD and IndL for ASC-ASC incremental learning (Task 0 to Task 1 of Table \ref{tab:ASC-ASC-AT}). Red lines separate the regions of new and old classes.}
    \label{fig:ASC-ASC}
    \vspace{-8pt}
\end{figure}

The experimental results  for the ASC-ASC-AT scenario are presented in Table~\ref{tab:ASC-ASC-AT} and demonstrate the effectiveness of independent learning of acoustic scenes (as $\mathbf{o}^{old}, \mathbf{o}^{new}$) from tasks 0 and 1. Note that ASC-ASC tasks use different audio recordings, in contrast to ASC-AT using the same audio recordings with different labels.

As an ablation study, we experiment with a few combinations of LR, using the same and reduced LR in incremental tasks. Results show that the use of the same LR (0.1, 0.01, or 0.001) in initial and incremental time steps makes the learner either fail to learn new acoustic scene classes effectively (showing high-stability) or significantly forget the old acoustic scene classes (showing high-plasticity). This situation is also known as the stability-plasticity dilemma of a learner between new and old knowledge. 
Using a LR of 0.1 or 0.01 in all the steps does not affect the performance of AT task; LR 0.001 seems not suitable to solve either task, while 0.01 seems to be best for the initial ASC.
Based on the results in Table~\ref{tab:ASC-ASC-AT}, the combination using a LR of 0.1 and 0.01  provide a best balance between stability and plasticity, with a similar performance on the two ASC tasks at both $t=1$ and $t=2$.

Looking at learning with or without IndL, we observe a large difference in forgetting between tasks. A learner without IndL learns from Eq.~(\ref{eq1}) using all logits $\mathbf{o}$ (combination of $\mathbf{o}^{old}, \mathbf{o}^{new}$) and class labels $\mathbf{y}^{\lctau_1}$. 
Because when learning the incremental ASC task 1, the learner does not have access to the data $(\mathbf{x}^{\lctau_0}, \mathbf{y}^{\lctau_0})$ of previous ASC task  0, the values of the task 0 targets in $\mathbf{y}^{\lctau_1}$ are zero. This makes the learner forget the old acoustic scene classes because it sees no examples of them. Hence, the amount of forgetting reached 65.2 p.p. and accuracy dropped to almost zero on the acoustic scene classes of task 0 at time step 1 (last row in Table~\ref{tab:ASC-ASC-AT}).  
In contrast, IndL of ASC task 1 with KD %from the previous task %does not much disturb the weights of the ASC task 0 and 
achieves an accuracy of 54.6\%  on the acoustic scene classes of task 0, with only a 10.7 p.p. forgetting. 
The two cases are illustrated in Fig.~\ref{fig:ASC-ASC}: without KD and IndL the network mostly predicts new classes (Fig.~\ref{fig:ASC-ASC}b), while using KD and IndL rebalances the output (Fig.~\ref{fig:ASC-ASC}a). 

It is worth noting that the F1 score of the AT task at $t=2$ is unaffected and it remains at 53.0\%, the same as the individual AT system. This is because the AT task is always learned independently of acoustic scenes due to a different loss function, and takes advantage of a previous model trained on highly relevant acoustic material, even though the class information differs. 
This suggests that an independent learning mechanism is a suitable approach for learning all the tasks.
\section{Conclusions and future work}

In this paper, we presented incremental ASC-AT and ASC-ASC-AT systems to solve distinct tasks over time. Results show that the performance of the ASC-AT system is close to the individual ASC and AT systems and outperforms joint ASC-AT learning with a similar size architecture. 
Independent learning of previous and current tasks with knowledge distillation significantly decreased the problem of catastrophic forgetting.
In the presented setup, the AT task is always independent of the ASC task; hence, learning these tasks would not much disturb the performances of one another irrespective of their order (whether ASC-AT or AT-ASC). 
Future work includes more detailed ablation studies of the different choices used to reduce forgetting and to improve overall performance, such as the order, size and type of the incremental tasks, and the use of cosine normalization.

\bibliographystyle{IEEEtran}
\bibliography{references}

\end{sloppy}
\end{document}